\documentclass[aps,prb,showpacs,raggedbottom,nobalancelastpage,amssymb,twocolumn,groupedaddress]{revtex4}
\usepackage{graphicx}
\usepackage{amsmath}
\usepackage{amsfonts}
\usepackage{amssymb}
\usepackage{epsfig}
\usepackage{color}
\usepackage{dsfont}

\newcommand{\abs}[1]{\left\vert#1\right\vert}

\newcommand{\bra}[1]{\left\langle#1\right\vert}
\newcommand{\ket}[1]{\left\vert#1\right\rangle}

\begin{document}

\title{Phase Diagram and Quench Dynamics in a Spinful Interacting Kitaev Chain}

\author{G.~Francica}
\affiliation{CNR-SPIN, I-84084 Fisciano (Salerno), Italy}

\author{P.~Gentile}
\affiliation{CNR-SPIN, I-84084 Fisciano (Salerno), Italy}\affiliation{Dipartimento di Fisica ``E. R. Caianiello'', Universit\`a di Salerno, I-84084 Fisciano (Salerno), Italy}

\author{M.~Cuoco}
\affiliation{CNR-SPIN, I-84084 Fisciano (Salerno), Italy}\affiliation{Dipartimento di Fisica ``E. R. Caianiello'', Universit\`a di Salerno, I-84084 Fisciano (Salerno), Italy}

\date{\today}

\begin{abstract}
We consider an exact solvable interacting spinful Kitaev chain which is a generalization of the Mattis-Nam model. A nearest-neighbor dimerized interaction favoring the production of disjoint molecules drives the quantum phase into an insulating one. The phases are characterized statically and dynamically in terms of  magnetization and spin-singlet correlations by using the exact solution.
The model is shown to be exactly solvable also in the presence of boundary interactions which are originated from a spin-singlet superconducting pairing and a magnetic field. We exploit the exact solution to investigate the out-of-equilibrium dynamics as due to a quench at the boundary. The propagation of the disturbance in nearest-neighbor magnetic and spin-singlet pairing displays a ballistic behavior for long times with different velocities.
\end{abstract}

\maketitle

\section{\label{sec.intro}Introduction}

In the last decades topological materials have received a great attention. Although a systematic study can be performed for non-interacting fermions~\cite{schnyder08,kitaev08,chiu16}, the role of interactions remains particularly attractive because it leads to a breakdown of this classification~\cite{Fidowski10}.

Unfortunately, interacting models which can be exactly solvable are uncommon. For example, one dimensional models like the Hubbard and Heisenberg models and some of their generalizations can be solved through a Bethe ansatz technique~\cite{takashi,korepin,kawakami}.

The Kitaev chain model in the presence of nearest neighbor interaction has been recently characterized by considering a frustration-free case~\cite{katsura15} with a non-homogenous chemical potential~\cite{Wouters18}, and an exact solution has been given at the symmetric point~\cite{Miao17} by also including dimerization~\cite{Ezawa17,Wang17}.
Another exactly solvable model which is a p-wave chain of spin $\frac{1}{2}$ fermions in presence of Hubbard interaction has been introduced by Mattis and Nam~\cite{Mattis72}. That model can be mapped onto two interacting Kitaev chains, which have been recently taken in exam outside the half-filling by adopting numerical methods and bosonization technics~\cite{Herviou16}.

A common aspect of these models is the presence of a topological non trivial ground state and the potential to describe a topological transition due to the interaction. In this context, the study of out of equilibrium dynamics of many-body systems following a quench, i.e an abrupt change of some control parameter of the system, can contain rich effects as it is typically related to the topological features of the achieved quantum phases. In particular, in one dimensional topological systems of non-interacting fermions, quench between distinct topologically phases are intimately related to the emergence of dynamical phase transitions~\cite{heyl18}. Moreover, the dynamics following a local quench allows to study the propagation of quasiparticles in the system, which has been already analysed in a one dimensional quantum Ising chain~\cite{smacchia12,francica16}.

In this paper, starting from the Mattis-Nam model we consider the role of interacting terms that favor the generation of effective molecular-like configurations in the ground state. An exact solution in terms of non-interacting spinless fermions can be achieved by performing two Jordan-Wigner transformations.
We provide a characterization of the quantum phase in the bulk and at the boundary of the chain, highlighting the presence of long range order in the trivial phases related to a spontaneous symmetry breaking of the $Z_2^x$ symmetry corresponding to a rotation by $\pi$ with respect to the $x$ axis in the spin space.  We demonstrate that the model is exactly solvable also in the presence of boundary interactions originating from a spin-singlet superconducting pairing and a magnetic field along a specific direction.
We also analyse the dynamics in the Mattis-Nam model due to a boundary quench by examining the propagation of disturbance in the nearest-neighbor magnetization and spin-singlet correlations. Both display a ballistic long time propagation  with sensitively  different velocities.

The paper is structured in the following way. In the Sect.~\ref{sec.model} we introduce the model summarizing its main symmetries. The Sect.~\ref{sec.exactsol} is devoted to the exact solution of the model. In Sec.~\ref{sec.phasesquench} we characterize the quantum phases from a topological point of view and in terms of the ground-state correlations. We also give a dynamically characterization by performing a quench at the boundary of the chain. Sec.~\ref{sec.conclusions} is devoted to the summary of the results achieved and conclusions.

\section{\label{sec.model} The Model}
We consider a one-dimensional chain of spin $\frac{1}{2}$ fermions described by the Hamiltonian $H=H_0+H_1 + H_{edge}$.

For a chain  of length $L$  with open boundary conditions, the Hamiltonian $H_0$  reads

\begin{eqnarray*}
H_0 &=& \sum_{j=1,\sigma}^{L-1}  \left[-t (c^\dagger_{j\sigma} c_{j+1\sigma} + h.c.) - \Delta_\sigma  (c^\dagger_{j\sigma} c^\dagger_{j+1 \sigma} + h.c.)\right] \\
&& +  U \sum_{j=1}^L (2 n_{j \uparrow} -1)(2 n_{j\downarrow}-1)- \mu \sum_{j=1,\sigma}^L  (n_{j\sigma} -\frac{1}{2})
\end{eqnarray*}

\noindent describing an electronic system with nearest neighbor hopping, spin-triplet p-wave pairing and on-site Hubbard type interaction. The operators $c_{j\sigma}$ ($c^\dagger_{j\sigma}$) annihilates (creates) a fermion on site $j$ with spin $\sigma=\uparrow,\downarrow$ and satisfy the anticommutation relations $\{c_{i\sigma},c_{j\sigma'}\}=0$ and $\{c^\dagger_{i\sigma},c_{j\sigma'}\}=\delta_{ij}\delta_{\sigma \sigma'}$, $n_{j\sigma}=c^\dagger_{j\sigma}c_{j\sigma}$ is the occupation number operator, $t$ is the hopping amplitude, $\Delta_\sigma$ is the superconducting pairing potential, $\mu$ is the chemical potential and $U$ is the interaction.
In particular, at the symmetric point $\Delta_\sigma=t$ and for $\mu=0$ the Hamiltonian $H_0$ reduces to the Mattis-Nam model.

The term $H_1$ reads

\begin{eqnarray}
\nonumber H_1 &=& \sum_{j=1}^{\frac{L}{2}} \lambda_j (c_{2j\uparrow}+c^\dagger_{2j \uparrow}) (c_{2j\downarrow}+c^\dagger_{2j \downarrow})(c_{2j-1\uparrow}-c^\dagger_{2j-1\uparrow})\\
&& \times (c^\dagger_{2j-1\downarrow}-c_{2j-1\downarrow})
\end{eqnarray}

\noindent and gives a dimerized nearest-neighbor interaction, where $\lambda_j$ is the coupling between sites $2j$ and $2j-1$, allowing the production of effective molecular like configurations in the ground state of the model.

We take into account boundary interactions described by $H_{edge}$, originating from the presence of a singlet superconductive pairing $\Delta_0$ and a magnetic field $\mathbf h=(h_x,h_y,h_z)$

\begin{equation}
H_{edge} = - \mathbf h \cdot \mathbf S_L -i\Delta_0 (c_{L \uparrow}c_{L \downarrow}-h.c)
\end{equation}

\noindent where $\mathbf S_j = (S_j^x,S_j^y,S_j^z)$ with $S^\alpha_j = \frac{1}{2}\sum_{\sigma \sigma'}c^\dagger_{j\sigma} \sigma^\alpha_{\sigma\sigma'}c_{j \sigma'}$ and $\{\sigma^\alpha\}$ are the Pauli matrices.

We assume an equal spin pairing triplet pairing, such that the order parameter has the same amplitude in both spin up and down channels $\Delta_\uparrow=\Delta_\downarrow=\Delta$.

Moroever, one can exploit the transformation $(c_{j \uparrow}, c_{j \downarrow})\to ( c_{j \uparrow}, i c_{j \downarrow})$ to explore the case of a spin triplet superconductor having opposite sign between the spin up and down triplet pairing. In terms of the $d$-vector representation, it implies that one can both investigate the case with $d_x$ and $d_y$ spin triplet pairing. The interaction transforms as $H_1\mapsto \sum_j \lambda_j (2 n_{2j \uparrow} -1)(2 n_{2j-1\downarrow}-1) $, i.e. a dimerized nearest-neighbor Hubbard interaction, and the boundary couplings as $\mathbf h \mapsto (h_y,h_x,h_z)$ and $\Delta_0 \mapsto i \Delta_0$.

We briefly recall the symmetry features of the model without the boundary term $H_{edge}$.

The model is symmetric with respect to the time-reversal transformation which is represented by an anti-unitary operator which acts on the fermions as $(c_{j \uparrow}, c_{j \downarrow})\to ( c_{j \downarrow},- c_{j \uparrow})$ and with respect to the number parity transformation represented by the unitary operator $Z^f_2 = e^{i \pi \sum n_{j\sigma}}$.
The model is invariant also under the rotation $(c_{j \uparrow}, c_{j \downarrow})\to ( c_{j \downarrow}, c_{j \uparrow})$ represented through the unitary operator $Z^x_2$.

At $\Delta=0$ there is a spin rotation symmetry $SU(2)$, which is broken down to the rotation around $y$ axis
\begin{equation*}
\left(
  \begin{array}{c}
    c_{j \uparrow} \\
    c_{j \downarrow} \\
  \end{array}
\right) \to \left(
            \begin{array}{cc}
              \cos \phi & -\sin \phi \\
              \sin \phi & \cos \phi \\
            \end{array}
          \right) \left(
  \begin{array}{c}
    c_{j \uparrow} \\
    c_{j \downarrow} \\
  \end{array}
\right)
\end{equation*}

\noindent due to the presence of the superconducting pairing. The associated conserved charge is the total spin in $y$ direction $S^{y} = \sum_{j=1}^L S^y_j$.


At $\mu=0$ the model is invariant under the particle-hole transformation $c_{i \sigma} \to (-1)^i c^\dagger_{i \sigma}$, so that $Z^p_2 = \prod_{j \sigma}(c_{j\sigma} + (-1)^j c^\dagger_{j\sigma})$ is conserved. We note that all the operators $Z^\gamma_2$ (with $\gamma=f,p,x$) squares to the identity (we consider  $L$ even).

Conversely, both the boundary interaction terms always break the time-reversal and the rotation $Z^x_2$ symmetries.

Because of the presence of the quartic terms, the Hamiltonian $H$ cannot be diagonalized in straightforward way.

In this paper we consider the symmetric point $\Delta = t$ at half-filling $\mu=0$, a magnetic field $\mathbf h= h \hat{\mathbf y}$ and $\lambda_j=\lambda$ homogenous.

\section{\label{sec.exactsol} Exact Solution}

We rewrite the Hamiltonian in the Majorana fermion representation by defining the real Majorana operators $a_{ j\sigma} = c_{ j\sigma} + c^\dagger_{ j\sigma} $ and $b_{ j\sigma} = -i c_{ j\sigma} + i c^\dagger_{ j\sigma}$ which satisfy relations $\{a_{i \sigma},a_{j \sigma'}\}= \{b_{i \sigma},b_{j \sigma'}\}=2 \delta_{ij}\delta_{\sigma \sigma'}$ and $\{a_{i \sigma},b_{j \sigma'}\}= 0$.

In this representation the Hamiltonian $H_0$ reads

\begin{equation}
 H_0 =  -i t\sum_{j =1,\sigma}^{L-1} a_{ j+1\sigma} b_{ j\sigma}  - U \sum_{j =1}^L  a_{ j\uparrow}b_{ j\uparrow} a_{ j\downarrow} b_{ j\downarrow}
\end{equation}

\noindent displaying decoupled Majorana fermions $\{a_{1\sigma},b_{ L\sigma}\}$ at $U=0$, and the interaction $H_1$ reads:

\begin{equation}
H_1 = \sum_{j=1}^{\frac{L}{2}} \lambda a_{2j\uparrow} a_{2j\downarrow}b_{2j-1\uparrow} b_{2j-1\downarrow}
\end{equation}

\noindent from which we  expect a trivial phase for strong interactions.

By performing two Jordan-Wigner transformations, we map the Hamiltonian $H$ onto the simpler fermionic model~\footnote{See appendix for details.}

\begin{eqnarray}\label{eq.Hami.K}
 H &=& - i U \sum_{j=1}^L a_j b_j - i t \sum_{j=1}^{L-1} (R_j+R_{j+1})a_j b_{j+1} \\
\nonumber && + \lambda\sum_{j=1}^{\frac{L}{2}} R_{2j-1}R_{2j}-\frac{i}{2}\left(\Delta_0 +\frac{h}{2}\right) R_L a_L b_L+c_0 R_L
\end{eqnarray}

\noindent which allows us to identify the constant of motion $R_j= i c_j d_j$, where $a_j$,$b_j$,$c_j$ and $d_j$ are real fermions which satisfy anticommutation relations $\{\alpha_{i},\beta_j\}=2 \delta_{\alpha \beta}\delta_{ij}$ (with $\alpha,\beta=a,b,c,d$) and we have defined the constant  $c_0= \frac{1}{2}\left(\Delta_0 - \frac{h}{2}\right)$.

The parities map onto $Z_2^f = \prod_j^{L}(i a_j b_j)$ and $Z_2^p = \prod_j^{L} (ic_jd_j)$.

All the eigenstates of the interacting model can be classified in terms of the eigenvalues $r_j$ of the operators  $R_j$, and the model can be exactly solvable in each of these subspaces. In details for the configuration $\{r_j\}$ we obtain a quadratic fermionic Hamiltonian $H = \frac{i}{2} \sum a_i B_{i j} b_j + c_0 r_L + \lambda \sum_{i=1}^{\frac{L}{2}}r_{2i}r_{2i-1}$ with $\mathbf B$ a real matrix with non zero elements  $B_{i,i+1} = -2(r_i + r_{i+1}) t$, $B_{ii}= -2 U$ for $i<L$ and $B_{LL}=-2\alpha U$ with $\alpha$ defined such that $\alpha U = \frac{1}{2}(\Delta_0+h/2)r_L + U$. By performing the singular value decomposition $\mathbf B = \mathbf U  \boldsymbol\Lambda  \mathbf V^T$ with $\mathbf U$ and $\mathbf V$ real orthogonal matrices and $\boldsymbol\Lambda $  semi-definite positive diagonal matrix with diagonal elements $\Lambda_k\geq 0$, the Hamiltonian can be written in the canonical form $H= \frac{i}{2}\sum_k \Lambda_k \tilde{a}_k \tilde{b}_k$ where $\tilde{a}_k = \sum_i U_{i k} a_i$ and $\tilde{b}_k = \sum_i V_{i k} b_i$ and depend on the configuration $\{r_j\}$.
The minimum energy in that sector results to be $E=-\sum_k \Lambda_k/2 + c_0 r_L + \lambda \sum_i r_{2i}r_{2i-1} $.
By focusing on the model without the boundary term $H_{edge}$, the two configurations $\{r_j\}$ and $\{-r_j\}$ give same energy spectrum due to the $Z^x_2$ symmetry.

The Mattis-Nam Hamiltonian $H_0$ has been previously characterized in ref.~\cite{Mattis72}, and in the following we clarify how the term $H_1$ changes the phase of the model.

A non-homogenous $r_j$ produces disjoint molecules (i.e. Kitaev chains) in Eq.~\eqref{eq.Hami.K} and for an attractive interaction $H_1$ ($\lambda<0$) the ground state energy is obtained for the two homogenous cases $\{r_i=1\}$ and $\{r_i=-1\}$ due to the lowering in energy coming from the molecules union.

Conversely, for $\lambda>0$ there  is a lowering in energy  as the configuration $\{r_i\}$ becomes non-homogeneous due to the $H_1$ term which gets  its minimum value $-\lambda L/2$ for the two staggered cases $\{r_{2j-1}=(-1)^{j+1},r_{2j}=(-1)^j\}$ and $\{r_{2j-1}=(-1)^{j},r_{2j}=(-1)^{j+1}\}$.

\begin{figure}
[h!]
\includegraphics[width=0.76\columnwidth]{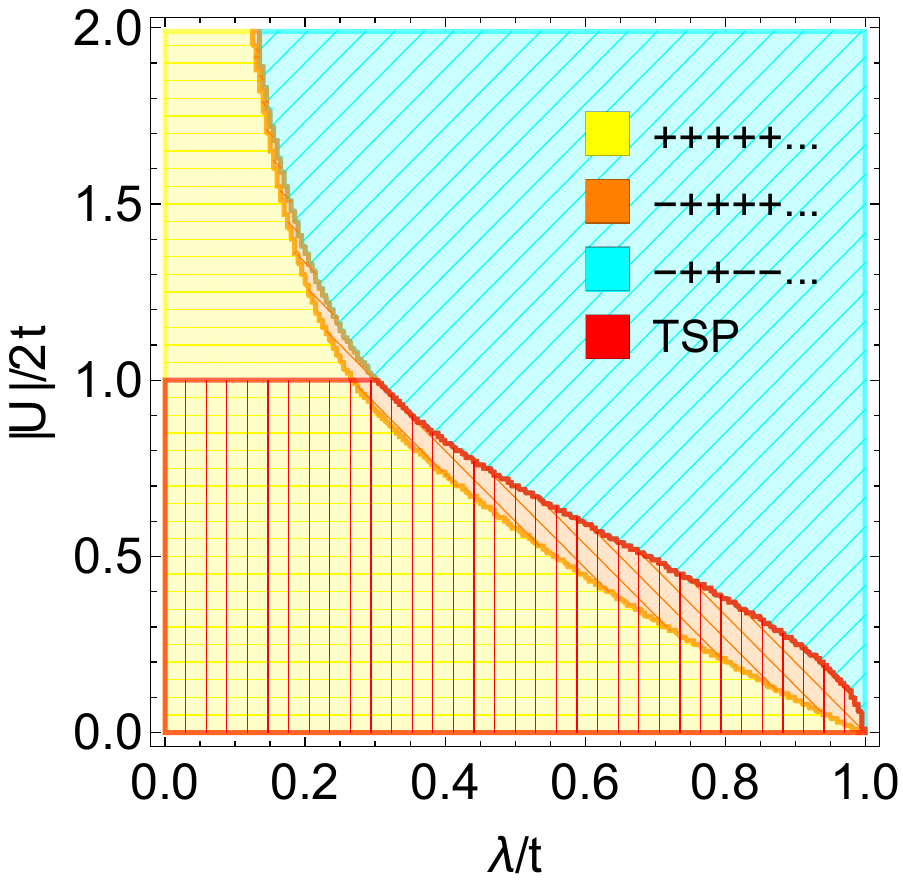}
\caption{ The phase diagram of the model in function of the interactions $U$ and $\lambda$. The yellow, orange and cyan regions correspond to the homogenous everywhere, only in bulk and staggered $r_j$, respectively. The region marked with red lines is topologically non-trivial. In the cyan region the phase is insulating, and the trivial yellow and orange regions show spin-singlet correlations for $U<-2t$ and anti-ferromagnetic order for $U>2t$.
We consider $L=16$ and we change $\lambda$ by steps of $0.005 t$ and $U$ by steps of $0.02 t$. 
}
\label{fig:lambda}
\end{figure}

For $U=0$, the ground state energy is obtained for the homogenous configurations if  $\lambda<t$ and  for the staggered ones  if $\lambda>t$. For $U\neq0$ a numerical analysis shows that the ground state energy  is obtained also for $r_j$ homogenous everywhere except the edges, i.e. homogenous only for $1<j<L$ (see Fig.~\ref{fig:lambda}).

Then in the bulk we have two distinct ground state sectors corresponding to the staggered (cyan region in Fig. 1) and homogenous configurations (yellow and orange regions in Fig. 1) due to the presence of the interaction $H_1$.

\section{\label{sec.phasesquench} quantum phases and quench}

In the two sectors $\{r_i=1\}$ and $\{r_i=-1\}$ the model with boundary interactions reads

\begin{eqnarray}\label{eq.Hami.K.1}
 \nonumber H &=& - i U \sum_{j=1}^L a_j b_j \mp i 2 t \sum_{j=1}^{L-1}a_j b_{j+1} \mp \frac{i}{2}\left(\Delta_0 +\frac{h}{2}\right) a_L b_L \\
 && \mp c_0 + \frac{\lambda L}{2}
\end{eqnarray}

\noindent and can be solved exactly in the limit $L\to \infty$ (the expressions of $U_{i k \pm}$, $V_{i k\pm}$ and $\Lambda_k$ can be read from the table in ref.~\cite{francica16}). In particular only for  $\abs{U}<2t$ there is one value $\Lambda_1 \sim 2 t \abs{\alpha}\frac{1-(U/2t)^2}{\sqrt{1+(\alpha^2-1)(U/2t)^2}}(U/2t)^L$ which vanishes as $L\to\infty$ with $U_{i 1\pm}$ and $V_{i 1\pm}$ exponentially localized at the sites $1$ and $L$. Except another possible out-band value, the other values  form a band $\Lambda(\theta)= 2 t\sqrt{(U/2t)^2+1-U/t\cos\theta}$ (for $U\neq 0$) in the thermodynamic limit.

For $c_0=0$, for every value of $U$ there are two ground states $\ket{0_\pm}$ belonging to the two subspaces $\{r_j =\pm1\}$ defined as the vacuum states such that $i\tilde a_{k \pm} \tilde b_{k\pm }\ket{0_\pm}= -\ket{0_\pm}$ for every $k$ and characterized by an even fermionic parity $\bra{0_\pm} Z^f_2\ket{0_\pm} = 1$. We observe that the two states are related via the transformation $Z^x_2$. This degeneration will be lifted due the presence of the term $H_{edge}$ for $c_0\neq 0$, and their energy will be $E_{\pm} = -\frac{1}{2}\sum_k \Lambda_k \pm c_0 +\lambda L/2$.

We note that the symmetry $Z^p_2$ can be expressed as the product of the two operators $Z^p_2 = \prod_\sigma Z^p_{2,\sigma}$ which shows the critical behavior

\begin{equation}
\lim_{L\to\infty}\abs{\bra{0_\pm} Z^p_{2,\sigma}\ket{0_\pm} }\sim \left\{
                                                    \begin{array}{ll}
                                                     1-\left(\frac{U}{2t}\right)^{2}, & \abs{U}<2t \\
                                                      O(1/L), & \abs{U}>2t
                                                    \end{array}
                                                  \right.
\end{equation}

\noindent For $\abs{U}<2t$, the two states $\ket{1_\pm} = \frac{1}{2}(\tilde a_{1 \pm} -i \tilde b_{1 \pm })\ket{0_\pm}$ having odd fermionic parity $\bra{1_\pm} Z^f_2\ket{1_\pm} = -1$, are degenerate for $c_0=0$, so that the ground state is fourfold degenerate in the limit $L\to\infty$.

The degeneration at $c_0=0$ is due to the presence of the symmetry under the composition $Z^x_2 T$, where $T$ is the time reversal transformation.

Without the edge term $H_{edge}$, we recall that in absence of interactions ($U=0$ and $\lambda=0$) the quantum phase exhibits a doublet of Majorana fermions at each edge. Then, in the presence of interactions we have a topological phase in the adiabatically connected red region, i.e. until we reach the phase transition line $U=\pm 2 t$ and $\lambda<\lambda_c$ at which the model is gapless, or until we reach a level crossing and the configuration $r_j$ becomes staggered.

We argue that the topological phase belongs to the class of topologically symmetry protected (TSP) states~\cite{Fidowski11}. We recall that the model is characterized through a time reversal symmetry represented by the antiunitary operator $T$, such that $T^2=-1$, and the fermionic parity $Z^f_2$, commutating each other. At the left boundary of the chain for $U=\lambda=0$ we consider the Hilbert space of the Majorana fermions $a_{1\uparrow}$, $a_{1\downarrow}$. By defining the complex fermion $c=\frac{1}{2}(a_{1\uparrow}+ia_{1\downarrow})$, the transformations $Z^f_2$ and $T$ are locally represented through $\hat Z^f_2 = (-1)^{c^\dagger c}$ and $\hat T = (c-ic^\dagger) K$ ($K$ being the complex conjugate operator), which implies that $\hat T^4=-1$ and $\hat Z^f_2 \hat T = - \hat T \hat Z^f_2$ until the ground state remains gapped.

Instead, in the two sectors $\{r_{2j-1}=(-1)^{j+1},r_{2j}=(-1)^j\}$ and $\{r_{2j-1}=(-1)^{j},r_{2j}=(-1)^{j+1}\}$ we have a sum of disjoint molecules

\begin{eqnarray}
\nonumber H &=& -iU (a_1b_1 + a_Lb_L) + \sum_{j=1}^{\frac{L}{2}} H^{(m)}_j \mp \frac{i}{2}\left(\Delta_0 +\frac{h}{2}\right) a_L b_L \\
 && \mp c_0 - \frac{\lambda L}{2}
\end{eqnarray}

\noindent where  $H_j^{(m)} = -iU (a_{2j}b_{2j} + a_{2j+1}b_{2j+1}) \mp i (-1)^j 2 t a_{2j} b_{2j+1}$ describes a molecule of length two. In this case the quantum phase (cyan region) is insulating and topologically trivial.

In the following we briefly summarize the main features of the quantum phases, by considering the thermodynamic limit $L\to \infty$.

We start by considering the homogenous cases $\{r_j=1\}$ and $\{r_j=-1\}$, i.e. the Hamiltonian $H_0$ is dominant over  $H_1$.

Local spin-singlet correlations tend to be formed for an attractive interaction and antiferromagnetic order for a repulsive one. For $\alpha=1$ the ground state shows spin-singlet correlations at the boundary, which are given by

\begin{equation}
\langle c_{L\uparrow} c_{L\downarrow} \rangle = i \frac{r_L}{4}\left( 1 - \frac{U}{\pi t}\int_0^\pi \frac{\sin^2(k)dk}{\sqrt{1+\left(\frac{U}{2 t}\right)^2 - \frac{U}{t} \cos(k)}} \right)
\end{equation}

\noindent It results that the correlation is non analytic at the critical points $U=\pm 2t$, and  for $U\ll - 2t $ two fermions tend to be localized at one edge $\langle c_{L\uparrow} c_{L\downarrow} \rangle\approx i r_L/2$, conversely for $U\gg 2t $ we have $\langle c_{L\uparrow} c_{L\downarrow} \rangle\approx 0$.

The boundary magnetization in the $y$ direction is related to the singlet correlation through the equation $\langle S^y_L \rangle  = -i\langle c_{L\uparrow} c_{L\downarrow} \rangle -\frac{r_L }{2} $, so that $\langle S^y_L \rangle \approx 0$ for $U\ll -2t$ and  $\langle S^y_L \rangle \approx \frac{r_L}{2}$ for $U\gg 2t$.
When the boundary interactions are such that $\alpha = 0$, in the correspondent subspace the real fermion $a_L$ is decoupled, and at finite size the boundary spin-singlet correlator tends to be
\begin{equation}
\lim_{\alpha\to 0} \langle c_{L\uparrow} c_{L \downarrow} \rangle = i\frac{r_L}{4}\left( 1-\frac{2t}{ U}\sqrt{\frac{\left(\frac{U}{2t}\right)^2-1}{1-\left(\frac{U}{2t}\right)^{-2L}}}\right)
\end{equation}

\noindent In order to characterize the quantum phase in the bulk, we will impose periodic boundary condition to the model in Eq.~\eqref{eq.Hami.K}.
Local spin-singlet correlation is generated only in the trivial phase $U<-2t$, and reads
\begin{equation}
\langle c_{i\uparrow} c_{i\downarrow} \rangle \sim \left\{
                                                    \begin{array}{ll}
                                                     i\frac{r_i}{2}\left(1-\left(\frac{2 t}{U}\right)^2\right)^{\frac{1}{4}}, & U<-2t \\
                                                      0, & U>-2t
                                                    \end{array}
                                                  \right.
\end{equation}

\noindent so that the phase exhibits a spontaneous breaking of the $Z_2^x$ symmetry and at long-range we have
\begin{equation}
\lim_{\abs{i-j}\to \infty} \langle  c_{i\uparrow} c_{i\downarrow}   c_{j\downarrow}^\dagger c^\dagger_{j\uparrow}  \rangle \sim \left\{
                                                    \begin{array}{ll}
                                                     \frac{1}{4} \left(1-\left(\frac{2 t}{U}\right)^2\right)^{\frac{1}{4}}, & U<-2t \\
                                                      0, & U>-2t
                                                    \end{array}
                                                  \right.
\end{equation}

\noindent Similarly, a spontaneous nonzero magnetization $\langle S_j^y\rangle$ is generated in the trivial phase $U>2t$ due to the repulsive interaction

\begin{equation}
\langle S_j^y \rangle \sim \left\{
                                                    \begin{array}{ll}
                                                     \frac{r_j (-1)^L(-1)^{j+1}}{2}\left(1-\left(\frac{2 t}{U}\right)^2\right)^{\frac{1}{4}}, & U>2t \\
                                                      0, & U<2t
                                                    \end{array}
                                                  \right.
\end{equation}

\noindent and the phase shows anti-ferromagnetic long-range order
\begin{equation}
\lim_{\abs{i-j}\to \infty}\langle S_i^y  S_j^y \rangle \sim \left\{
                                                    \begin{array}{ll}
                                                     \frac{(-1)^{j-i} }{4} \left(1-\left(\frac{2 t}{U}\right)^2\right)^{\frac{1}{4}}, & U>2t \\
                                                      0, & U<2t
                                                    \end{array}
                                                  \right.
\end{equation}

\noindent which vanishes exponentially for $U<2t$.

On the other side, when $H_1$ is dominant and the configuration $\{r_{2j}=(-1)^j,r_{2j-1}=(-1)^{j+1}\}$ are selected, the phase is characterized by the correlations

\begin{equation}
\langle  c_{i\uparrow} c_{i\downarrow}   c_{j\downarrow}^\dagger c^\dagger_{j\uparrow}   \rangle = (-1)^P\frac{r_i r_j}{8}\left( \frac{U}{\sqrt{t^2+U^2}}-1\right)
\end{equation}

\noindent and

\begin{equation}
\langle S_i^y  S_j^y \rangle = (-1)^P\frac{r_i r_j}{8}\left( \frac{U}{\sqrt{t^2+U^2}}+1\right)
\end{equation}

\noindent where $(-1)^P=1$ if $i$ and $j$ have same parity and $(-1)^P=-1$ otherwise. From which we deduce that the correlation of the spin state between two neighbor sites belonging to (two distinct) the same molecule is  (ferromagnetic) anti-ferromagnetic.

For this model we note that the exact solution allows us to explore which role a non-homogenous interaction $\underline U = (U_1,U_2,\cdots,U_L)$ plays, where $U_i$ is its value on the site $i$. As an example, we consider $\lambda=0$ and a pattern $\underline{U}=(x,\cdots,x,1,x,\cdots,x,1,x\cdots)U$ where the value $U$ is repeated in steps of $N$. From a numeric calculation of the Majorana number for the model in Eq.~\eqref{eq.Hami.K.1} it results that the phase is expected to be topologically nontrivial if $t> \beta_{N}(x) \abs{U}$ where the linear coefficient $\beta_{N}(x)$ decreases to zero as $x\to 0$ and to $x/2$ as $N\to L$. The same result is observed by considering random distributions of the interaction $U$ in the pattern $\underline{U}$, with a linear coefficient that decreases with the occurrence of the values $x$ in $\vec U$. This analysis proves how the topological phase is strong against a sparse interaction ($x=0$) in the chain.

We now proceed giving a characterization of the dynamics features.
By changing in time $\tau$ the on-site interaction $U$, we produce the power $\mathcal P = \langle \dot H (\tau) \rangle = \dot U \sum_{j=1}^L  \langle -i a_j(\tau) b_j(\tau) \rangle $. A particularly interesting case arises when considering the driving across the critical line $U=\pm 2t$ and $\lambda<\lambda_c$ which gives a nonzero power due to the production of adiabatic excitations related to the second order quantum phase transition.
As an example, we consider a turn-on of a strong interaction $U\gg 2 t$ which is linear in time with speed $\dot U= 1/\tau_Q$. Due to the relation to the Ising model, we have the power density  $\frac{\mathcal P}{L} \sim \frac{1}{2\pi} \frac{\dot U}{2t\sqrt{2\tau_Q }}$~\cite{Dziarmaga05}.

We now turn to the study of the dynamics in the non-insulating phase following the sudden turn-off of the boundary interactions.  With the aim to show how a local perturbation propagates in the interacting chain, we assume the system to be initially prepared in the ground state of $H$ (with $\alpha=0^+$). At $\tau = 0$ the boundary interaction is suddenly switched-off ($\alpha=1$ for $\tau>0$), so that the time evolution is generated by the Mattis-Nam Hamiltonian $H_0$. The dynamics remains in the invariant subspace with homogenous $r_j$, giving a ballistic propagation of the signal.

In particular, the density of particles is constant $\langle n_{j\sigma}\rangle = \frac{1}{2}$, so that the propagation occurs with a zero normal current.

\begin{figure}
[h!]
\includegraphics[width=0.49\columnwidth]{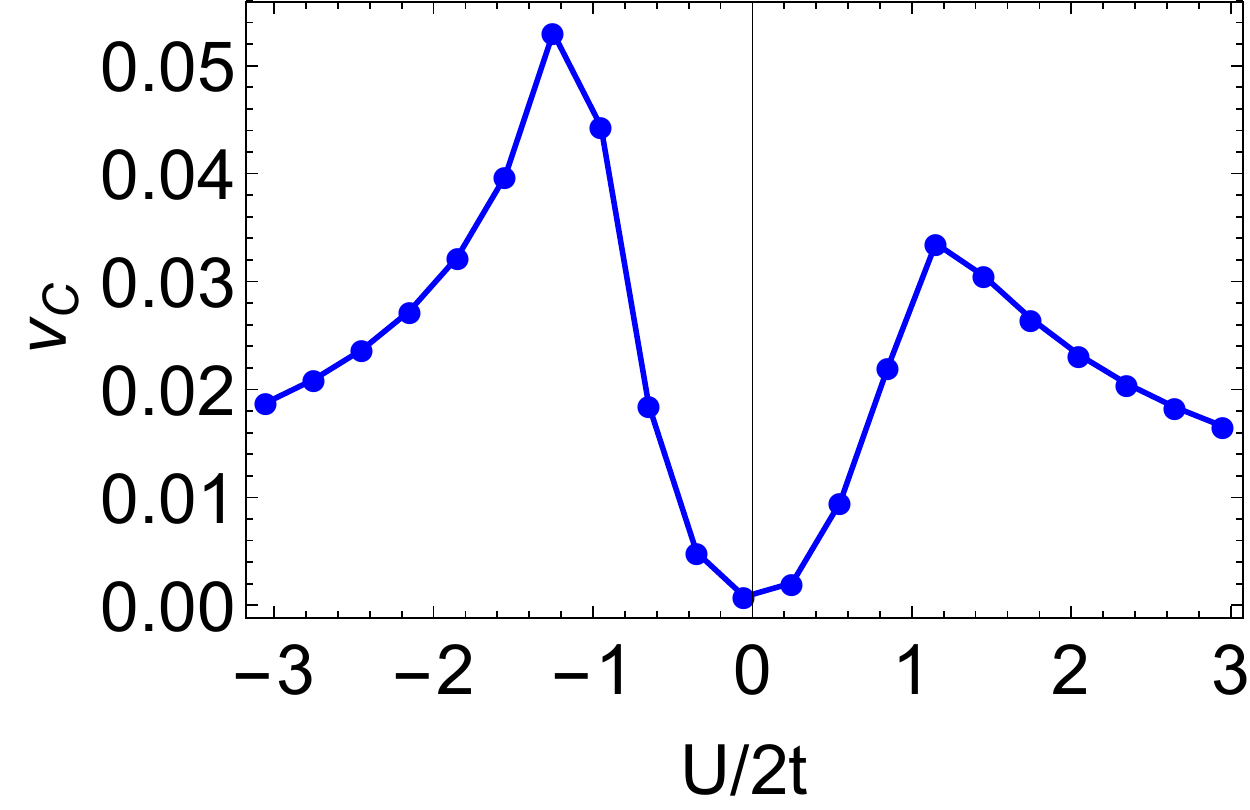} \includegraphics[width=0.49\columnwidth]{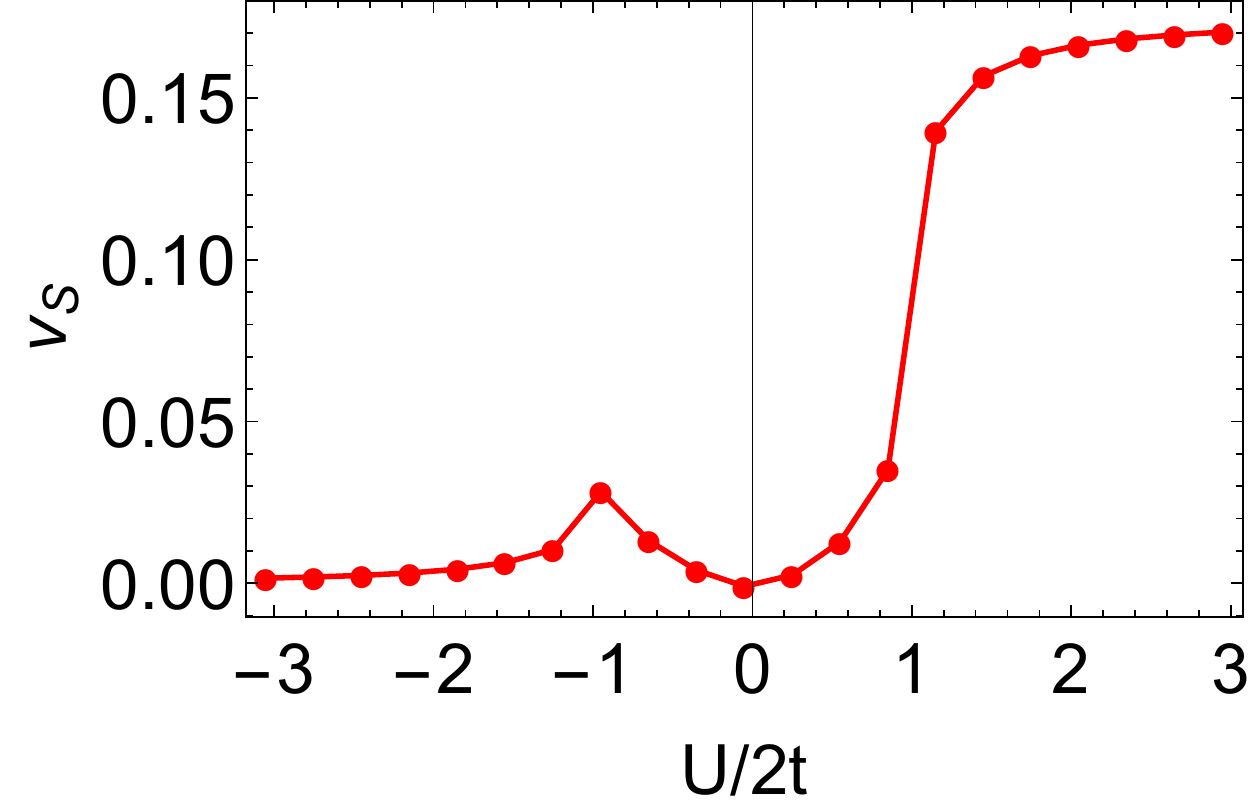}
\caption{The velocities $v_C(\tau)$ and $v_S(\tau)$ in the ballistic regime ($\tau=30$) in function of the interaction $U$. We turn-off the boundary term $H_{edge}$ from the initial value $\alpha = 0^+$. We put $L=200$.}
\label{fig:Rt}
\end{figure}

We consider how the nearest neighbor charge and spin correlations $C_i(\tau) = \langle c_{i \uparrow}(\tau) c_{i \downarrow}(\tau)c^\dagger_{i+1 \downarrow}(\tau) c_{i+1 \uparrow }^\dagger(\tau)\rangle $ and $S_i(\tau) = \langle S^y_{i}(\tau) S^y_{i+1}(\tau)\rangle $ evolve in time after a local sudden quench generated by changing the edge term at the initial time $\tau=0$.  By considering the mean square center $R_f^2(\tau) = (1/L)\sum_i f_i(\tau) (i-L)^2 $ where $f=C,S$, we numerically find that the velocity $v_f(\tau) = \partial_\tau \sqrt{\abs{R_f^2(\tau) - R_f^2(0)}}$ tends to be constant for $\tau\gg 1$ (see fig.~\ref{fig:Rt}), so that the propagation is ballistic.

We note that the difference in the propagation velocities $v_C$ and $v_S$ is due to how the modes interfere, producing a spin-charge separation effect.

\section{\label{sec.conclusions} conclusion}
In conclusion we have characterized the static and dynamical properties of the Mattis-Nam model in the presence of both nearest-neighbor interactions and magnetic and superconducting couplings at the boundary.
The model is shown to be exactly solvable by a mapping onto a non-interacting one, allowing to give a direct access to the characterization of the quantum phase diagram. Due to the new interaction an insulating phase emerges with an electronic pattern which is marked by effective disjoint molecular-like configurations. The mean features of the different phases have been characterized from a topological point of view, and by looking at the superconducting spin-singlet and magnetic correlations. In particular the topological phase exhibits four Majorana fermions in absence of quartic interactions, and by turning-on these interactions the model can undergo a transition into trivial phases. Remarkably, we find that the transition into the insulating region occurs via a distinct inhomogeneous intermediate phase showing edges disjoint from the bulk. We provide a characterization also from a dynamical point of view by considering the evolution due to a quench generated by suddenly removing the boundary term. The outcome highlights the possibility of obtaining a propagation as a function of the Hubbard interaction strength for the magnetic and superconducting correlators with significantly different velocities.

\appendix

\section{Mapping}
We rewrite the Hamiltonian $H_0$ in the Majorana fermion representation defined in the main-text 


\begin{eqnarray*}
 \nonumber H_0&=&  -\frac{i}{2} \sum_{j =1,\sigma}^{L-1}\left[(t+\Delta_\sigma) a_{ j+1\sigma} b_{ j\sigma}+(t-\Delta_\sigma) a_{ j\sigma} b_{ j+1\sigma} \right] \\
 &&  -U\sum_{j =1}^L a_{ j\uparrow}b_{ j\uparrow} a_{ j\downarrow} b_{ j\downarrow} -i\frac{\mu}{2} \sum_{j=1,\sigma}^{L}a_{ i\sigma} b_{ i\sigma}
\end{eqnarray*}

\noindent By performing the Jordan-Wigner transformation

\begin{eqnarray*}
a_{ j \uparrow} &=& \left(\prod_{i=1}^{j-1} \sigma^x_{2 i-1}\right)\sigma^z_{2 j-1}\\
b_{ j \uparrow} &=&  \left(\prod_{i=1}^{j-1} \sigma^x_{2 i-1}\right)\sigma^y_{2 j-1}\\
a_{ j \downarrow} &=&  \left(\prod_{i=1}^{L} \sigma^x_{2 i -1}\right) \left(\prod_{i=j+1}^{L} \sigma^x_{2 i}\right) \sigma^y_{2 j}\\
b_{ j \downarrow} &=& -\left(\prod_{i=1}^{L} \sigma^x_{2 i-1}\right) \left(\prod_{i=j+1}^{L} \sigma^x_{2 i}\right) \sigma^z_{2 j}
\end{eqnarray*}

\noindent the Hamiltonian is mapped onto the spin ladder model

\begin{eqnarray*}
H_0&=&   \sum_{j =1}^{L-1} \bigg(\frac{t+\Delta_\uparrow}{2}  \sigma^z_{2 j -1} \sigma^z_{2 j+1} +  \frac{t+\Delta_\downarrow}{2} \sigma^z_{ 2 j} \sigma^z_{2 j+2}   \\
 && + \frac{t-\Delta_\uparrow}{2}\sigma^y_{2 j-1} \sigma^y_{2 j+1} + \frac{t-\Delta_\downarrow}{2}\sigma^y_{2 j} \sigma^y_{2 j+2}  \bigg)\\
 &&  +  U\sum_{j =1}^L \sigma^x_{2j-1} \sigma^x_{2j} +\frac{\mu}{2}\sum_{j =1}^{2L} \sigma^x_{j}
\end{eqnarray*}

\noindent and  $Z_2^f = \prod_j^{2L} (-\sigma^x_j)$, $Z_2^p = \prod_j^{2L}( \sigma^z_j)$.

\begin{figure}
[h!]
\includegraphics[width=0.9\columnwidth]{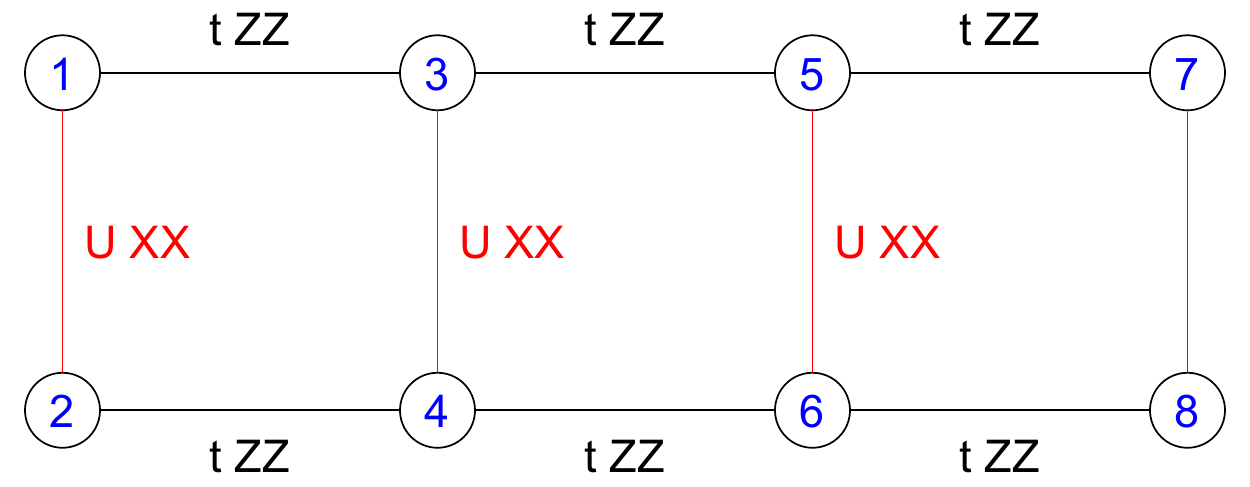}
\caption{ A representation of the spin ladder at the symmetry point $\Delta_\sigma=t$ and $\mu=0$ for $L=4$.
}
\end{figure}

\noindent We consider $\mu=0$, $t=\Delta_\uparrow=\Delta_\downarrow$ and without loss of generality $t$ is considered positive. The spin ladder model reduces to

\begin{equation*}
H_0 = U \sum_{j =1}^{L}  \sigma^x_{2j-1}\sigma^x_{2j}+  t \sum_{j =1}^{L-1}\left( \sigma^z_{2 j-1} \sigma^z_{2 j+1} +\sigma^z_{2 j} \sigma^z_{2 j+2} \right)
\end{equation*}

\noindent The mapping onto the spin ladder allow us to identify a set of observable $R_j= \sigma^z_{2j-1}\sigma^z_{2j}$ with $j=1,\cdots,L$ that commutate with the Hamiltonian~\cite{Brzezicki09}. The constant of motion $R_j$ has eigenvalue $r_j=\pm1$ and in terms of the fermionic operators reads $R_j = \left(\prod_{i=j}^L i b_{i\uparrow } b_{i \downarrow}\right) \left(\prod_{i=j+1}^L ia_{i\downarrow}a_{i \uparrow}\right)$.
The dependence on these observable is made explicit by recasting the Hamiltonian as

\begin{equation*}
H_0 =  U\sum_{j =1}^{L} \sigma^x_{2j-1}\sigma^x_{2j}+ t \sum_{j =1}^{L-1} \left( R_j + R_{j+1}\right)\sigma^z_{2 j} \sigma^z_{2 j+1}
\end{equation*}

\noindent By performing a second Jordan-Wigner

\begin{eqnarray*}
\alpha_j &=& \left( \prod_{i=1}^{j-1} \sigma^y_i\right) \sigma^x_j\\
\beta_j &=&  \left( \prod_{i=1}^{j-1} \sigma^y_i\right) \sigma^z_j
\end{eqnarray*}

\noindent we map the spin ladder onto the fermionic model

\begin{equation*}
H_0= - i U \sum_{j=1}^L a_j b_j - i t \sum_{j=1}^{L-1} (R_j+R_{j+1})a_j b_{j+1}
\end{equation*}

\noindent and the constant motion $R_j= i c_j d_j$, where we have renamed the Majorana operators $a_j=\alpha_{2j}$, $c_j=\beta_{2j}$,  $d_j=\alpha_{2j-1}$ , $b_j=\beta_{2j-1}$.

Conversely, the boundary term $ H_{edge}=-h S_L^y - \Delta_0 (i c_{L\uparrow}c_{L\downarrow} + h.c) $ is mapped onto

\begin{equation*}
H_{edge} =-\frac{i}{2}\left(\Delta_0 +\frac{h}{2}\right) R_L a_L b_L+\frac{1}{2}\left(\Delta_0 - \frac{h}{2}\right) R_L
\end{equation*}

\noindent and the dimerized interaction term $H_1$ onto

\begin{equation*}
H_1 = \sum_{j=1}^{\frac{L}{2}} \lambda_j R_{2j-1}R_{2j}
\end{equation*}

\noindent We note that the mapping  is the same of the mapping  in Ref.~\cite{Mattis72}.

\section{Correlations}

The spin-singlet correlation on the site $j$ can be expressed in terms of the expectation value of fermionic operators

\begin{equation*}
\langle c_{j\uparrow} c_{j\downarrow} \rangle = i\frac{r_j}{4}\left( \langle \prod_{i=j+1}^L (-i a_i b_i) \rangle +\langle  \prod_{i=j}^L (-i a_i b_i)  \rangle \right)
\end{equation*}

\noindent where the average is calculated with respect to an eigenstate with eigenvalues $\{r_j\}$.
At the same way we have

\begin{eqnarray*}
 \langle  c_{i\uparrow} c_{i\downarrow}   c_{j\downarrow}^\dagger c^\dagger_{j\uparrow}  \rangle &=& \frac{r_i r_j}{16} \bigg(\langle \prod_{l=i}^{j-1} (-i a_l b_l) \rangle + \langle \prod_{l=i+1}^{j-1} (-i a_l b_l) \rangle \\
 && +\langle \prod_{l=i}^{j} (-i a_l b_l) \rangle + \langle \prod_{l=i+1}^{j} (-i a_l b_l) \rangle \bigg)
\end{eqnarray*}

\noindent The magnetization in the $y$ direction can be calculated as

\begin{equation*}
\langle S_j^y \rangle = \frac{r_j}{4} \left( \langle \prod_{i=j}^L (-i a_i b_i) \rangle -\langle  \prod_{i=j+1}^L (-i a_i b_i)  \rangle \right)
\end{equation*}

\noindent and the spin correlations $\langle S_i^y  S_j^y \rangle$ as

\begin{eqnarray*}
 \langle S_i^y  S_j^y \rangle &=& \frac{r_i r_j}{16} \bigg(\langle \prod_{l=i}^{j-1} (-i a_l b_l) \rangle - \langle \prod_{l=i+1}^{j-1} (-i a_l b_l) \rangle \\
 && -\langle \prod_{l=i}^{j} (-i a_l b_l) \rangle + \langle \prod_{l=i+1}^{j} (-i a_l b_l) \rangle \bigg)
\end{eqnarray*}

\noindent The particular choice of the state gives $\langle c_{j \sigma} c_{j+1 \sigma'}\rangle = 0$ from which we note that nearest neighbor triplet or singlet correlations are not formed. There is not magnetization in the $x$ and $z$ directions $ \langle S_j^x \rangle  = \langle S_j^z \rangle = 0$, the density $\langle n_{j\sigma}\rangle=\frac{1}{2}$ and the correlation functions $\rho_{ij}^{\sigma\sigma'} = \langle (2 c^\dagger_{i \sigma}c_{i \sigma}-1)(2 c^\dagger_{j \sigma'}c_{j \sigma'}-1)\rangle$ are only local with $\rho_{jj}^{\uparrow\downarrow}=-i\langle a_j b_j\rangle$, from which $\langle S_i^z  S_j^z \rangle  = 0$.

Furthermore we have that

\begin{equation*}
Z_{2,\uparrow}^p = \prod_{j=1}^{\frac{L}{2}}(- r_{2j-1}) \langle a_1 ib_2 a_3 ib_4 \cdots a_{L-1} ib_L\rangle
\end{equation*}

\noindent For the ground state, all these expectation values can be calculated with the help of the Wick theorem, from which $\langle \prod_{l=i}^{j} (-i a_l b_l)\rangle$ can be written as the determinant

\begin{equation*}
 \langle \prod_{l=i}^{j} (-i a_l b_l) \rangle = \left\vert
                                                  \begin{array}{cccc}
                                                    G_{i\,i} & G_{i\,i+1} & \cdots & G_{i\,j} \\
                                                    \vdots & \ddots &  & \vdots \\
                                                    G_{j\,i} &   &  & G_{j\,j} \\
                                                  \end{array}
                                                \right\vert
\end{equation*}

\noindent The element matrix $G_{i\,j}=-i\langle a_i b_j\rangle$ can be calculated as $G_{i\,j}= \sum_k U_{i k} V_{j k}$.

For $r_j$ homogenous, by imposing periodic boundary conditions to the model in Eq.~\eqref{eq.Hami.K} we have

\begin{equation*}
\lim_{\abs{i-j}\to \infty}\langle\prod_{l=i}^{j-1} (-i a_l b_l) \rangle \sim \left\{
                                                    \begin{array}{ll}
                                                     \left(\frac{U}{\abs{U}}\right)^{j-i}\left(1-\left(\frac{2 t}{U}\right)^2\right)^{\frac{1}{4}}, & \abs{U}<2t \\
                                                      0, & \abs{U}>2t
                                                    \end{array}
                                                  \right.
\end{equation*}

\noindent For the staggered case $\{r_{2j-1}=(-1)^{j+1}, r_{2j}=(-1)^j\}$, the model can be solved exactly at  finite $L$. The matrix $G$ is the diagonal block matrix

\begin{equation*}
G = \left(
      \begin{array}{cccccc}
        \frac{U}{\abs{U}} &  &  &  &  &  \\
         & o &  &  &  &  \\
         &  & o^T &  &  &  \\
         &  &  & o &  &  \\
         &  &  &  & \ddots  &  \\
         &  &  &  &  & \frac{U}{\abs{U}} \\
      \end{array}
    \right)
\end{equation*}

\noindent where $o$ is the matrix

\begin{equation*}
o = \frac{1}{\sqrt{t^2+U^2}}\left(
                              \begin{array}{cc}
                                -U & t \\
                                -t & -U \\
                              \end{array}
                            \right)
\end{equation*}

\noindent from which
\begin{equation*}
\langle\prod_{l=i}^{j} (-i a_l b_l) \rangle = \left\{
                                                    \begin{array}{ll}
                                                     o_{11}, & j \,\text{odd},\, i \,\text{odd} \bigvee j\, \text{even}, i \,\text{even} \\
                                                      1, & j \,\text{odd}, i \,\text{even} \bigvee j \,\text{even}, i \,\text{odd}
                                                    \end{array}
                                                  \right.
\end{equation*}

\noindent for $i>1$ and $j<L$.

The values of $\Lambda$ are $\Lambda_{1,L}=2\abs{U}$ and $\Lambda_{2i,2i+1}=2\sqrt{2 t^2 +U^2 \pm 2 t\sqrt{t^2 + U^2}}$ for $i=1,\cdots,L/2-1$.

\subsection{Dynamics}

The study of the dynamics is simplified thanks to the mapping on the non-interacting model.

In the invariant subspace with eigenvalues $\{r_j\}$, in the Heisenberg picture we have the motion equations

\begin{eqnarray*}
\dot a_i(\tau) &=& \sum_j  B_{ij}(\tau) b_j(\tau)\\
\dot b_i(\tau) &=& - \sum_j B_{ji}(\tau) a_j(\tau)
\end{eqnarray*}

\noindent where the time is indicated with $\tau$.

The time evolved vector $\boldsymbol\psi = (a_1, \cdots, a_L,b_1,\cdots, b_L)^T$ can be expressed as $\boldsymbol\psi (\tau) = \boldsymbol \Phi(\tau)\boldsymbol\psi(0) $,  where $\boldsymbol\Phi(\tau)$ is solution of the differential equation $\dot {\boldsymbol\Phi}(\tau) = i \tau_y \otimes B(\tau) \boldsymbol\Phi(\tau)$, with initial condition $\boldsymbol\Phi(0)=1$, where $\tau_y$ is the Pauli matrix $\tau_y = \left(
                                \begin{array}{cc}
                                  0 & -i \\
                                  i & 0 \\
                                \end{array}
                              \right)$.

For a sudden quench where the initial state is the ground state of the Hamiltonian $H' = \frac{i}{2} \sum_{ij} a_i B'_{ij} b_j $, we have


\begin{eqnarray*}
 a_i(\tau) &=& \sum_{k} X_{11,ik}(\tau) \tilde{a}'_k + X_{12,ik}(\tau) \tilde{b}'_k\\
 b_i(\tau) &=& \sum_{k} X_{21,ik}(\tau) \tilde{a}'_k + X_{22,ik}(\tau) \tilde{b}'_k\\
\end{eqnarray*}

\noindent where we have defined the matrices $X_{11}= U \cos(\Lambda \tau) U^T U'$, $X_{12}= U \sin(\Lambda \tau) V^T V'$, $X_{21}= -V \sin(\Lambda \tau) U^T U'$ and $X_{22}= V \cos(\Lambda \tau) V^T V'$, and the orthogonal matrices $U'$ and $V'$ are such that $B'=U' \Lambda' {V'}^T$ and $\tilde{a}'_k = \sum_i U'_{ik} a_i$, $\tilde{b}'_k = \sum_i V'_{ik} b_i$.

By exploiting the Wick theorem, the correlations at equal time like as $C_i(\tau)$, can be expressed in terms of the two points correlations

\begin{eqnarray*}
\langle -i a_i(\tau) b_j(\tau) \rangle &=& (X_{11}X_{22}^T - X_{12}X_{21}^T)_{ij}\\
\langle -i a_i(\tau) a_j(\tau) \rangle &=& (X_{11}X_{12}^T - X_{12}X_{11}^T)_{ij}\\
\langle -i b_i(\tau) b_j(\tau) \rangle &=& (X_{21}X_{22}^T - X_{22}X_{21}^T)_{ij}
\end{eqnarray*}

\end{document}